\documentclass[twocolumn,showpacs,preprintnumbers,amsmath,amssymb,superscriptaddress]{revtex4}
\usepackage{graphicx}
\usepackage{dcolumn}
\usepackage{bm}
\usepackage{epsfig}
\usepackage{CJK}
\usepackage{bm}

\begin{document}

\title{Electrically Tunable Topological State in [111] Perovskite Materials
 with Antiferromagnetic Exchange Field}

\author{Qi-Feng Liang}
\affiliation{International Center for Materials Nanoarchitectornics (WPI-MANA)
National Institute for Materials Science, Tsukuba 305-0044, Japan}
\affiliation{Department of Physics, Shaoxing University, Shaoxing 312000, China}
\author{Long-Hua Wu}
\affiliation{International Center for Materials Nanoarchitectornics (WPI-MANA)
National Institute for Materials Science, Tsukuba 305-0044, Japan}
\author{Xiao Hu}
\affiliation{International Center for Materials Nanoarchitectornics (WPI-MANA)
National Institute for Materials Science, Tsukuba 305-0044, Japan}
\date{\today}

\begin{abstract}
A topological state with simultaneous nonzero Chern number and spin Chern number
is possible for electrons on honeycomb lattice based on band engineering by
staggered electric potential and antiferromagnetic
exchange field in presence of intrinsic spin-orbit coupling.
With first principles calculation we confirm that the scheme can be realized by
material modification in perovskite G-type antiferromagnetic insulators grown along [111] direction,
where d electrons hop on a single buckled honeycomb lattice. This material
is ideal for spintronics applications, since it provides a spin-polarized quantized edge
current, robust to both nonmagnetic and magnetic defects, with the spin polarization
tunable by inverting electric field.
\end{abstract}

\maketitle

\vspace{3mm} \textit{Introduction.---}The discovery of quantum Hall effect (QHE) by von Klitzing opened a new chapter in
condensed matter physics. It was revealed \cite{TKNN82,Niu85} that the integer coefficient
in Hall conductance is nothing but the Chern number labeling
the topological property of the electron wavefunction. Since then to explore
possible topological states has been one of the main driving forces in study of
condensed matter physics\cite{Haldane88,KaneMele05,BernevigScience06,KonigScience07,HasanRMP10,QiRMP11,XiaoNatComm11,XingPRL11}.

The breakthrough took place when Kane and Mele clarified that electrons
on graphene, a two-dimensional honeycomb lattice of carbon atoms, open a gap
by spin-orbit coupling (SOC) and achieve a topologically nontrivial
state called quantum spin Hall effect (QSHE) \cite{KaneMele05}.
It turned out that in graphene the intrinsic SOC is very small
which makes detection of QSHE hopeless.
QSHE was then predicted theoretically \cite{BernevigScience06} and realized experimentally in HgTe
quantum well with strong SOC \cite{KonigScience07}.

\begin{figure}
\psfig{figure=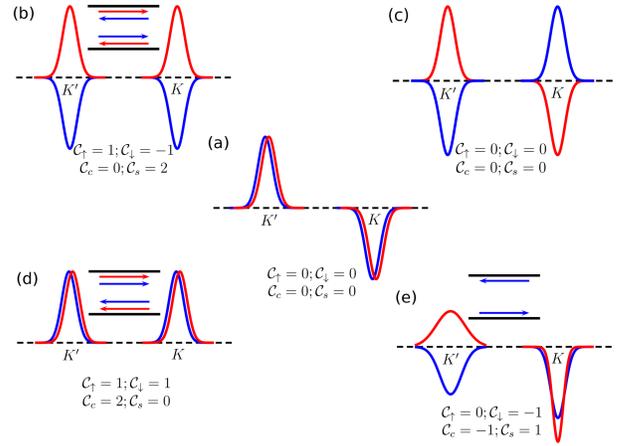,width=8cm}
\caption{(Colour online)
Schematics for possible configurations of Berry curvatures and Chern numbers for
insulating states on honeycomb lattice: (a) pristine honeycomb lattice,
(b) QSHE, (c) trivial state with broken time-reversal symmetry, (d) QAHE, (e) topological
state with simultaneous nonzero charge and spin Chern numbers.}
\label{fig:1}
\end{figure}

As the common homebase for both Kane-Mele model \cite{KaneMele05}and the spinless Haldane model \cite{Haldane88},
honeycomb lattice serves a unique role in understanding the topological property
of electron systems. It may be illustrative to summarize its electronic structure
paying attention to the Berry curvature configuration \cite{BerryCurvPRL05}.
There are two sites in the unit cell of honeycomb structure. With nearest neighbor hopping,
electronic valance and conduction bands touch linearly and thus form Dirac cones
at two inequivalent k points, K and K$'$, locating at the corners of Brillouin zone.
It is important to observe that Bloch wavefunctions exhibit opposite chiral features around K and K$'$,
characterized by opposite Berry curvatures, which establishes the special position of honeycomb lattice
in exploring topological state.

With time-reversal
symmetry, the Berry curvatures in spin-up and spin-down channels overlap as
shown in figure 1(a). Introducing intrinsic SOC opens a gap at the two Dirac cones,
and flips Berry curvatures at K in spin-up channel and at K$'$ in spin-down
channel preserving time-reversal symmetry, which results in the Berry curvature configuration in figure 1(b),
characterizing the QSHE state \cite{KaneMele05}.
One can further flip the Berry curvatures at K in both spin-up and spin-down channels, yielding
the configuration figure 1(c) (a topologically trivial state), by a field conjugate to
the valley degree of freedom such as the polarized field \cite{XiaoPRL07,ZengNatNano12,MakNatNano12,EZAWAPRL13}.
With an additional staggered electric field, which can be realized in silicene with
buckled honeycomb lattice in terms of a uniform electric field\cite{NiNanoLett11}, Ezawa could flip
the Berry curvature at K in only one spin channel, resulting in figure 1(e) characterized
by simultaneous nonzero charge Chern number and spin Chern number
\cite{EZAWAPRL13} (see also \cite{LiuCXPRL08,RueggPRB11,YangPRB11}).
The band engineering is based on a full
control on the degrees of freedom of spin, valley and sublattice, taking
advantages of the intrinsic SOC. By the way, the Berry curvature
configuration in figure 1(d) corresponds
to quantum anomalous Hall effect (QAHE)\cite{NagaosaRMP10,QiaoPRB10,YuScience10,RueggPRB11,YangPRB11,EZAWAPRL12}.

In the present paper we notice that in presence of intrinsic SOC, a staggered magnetic field plays
a similar role as provided by the polarized light proposed by Ezawa\cite{EZAWAPRL13}. Since the staggered magnetic
field can be realized by antiferromagnetic (AFM) insulators, compact and stable devices based
on the topological state in figure 1(e) are possible as compared with the photo-assisted scheme.

As material realization of our idea, we focus on d-electron systems in perovskite structure\cite{Mitchellbook}.
First, we choose a perovskite insulator ABO$_3$ with G-type AFM order on the
magnetic B atoms. As first discussed by Xiao et al.\cite{XiaoNatComm11}, along [111] direction
B atoms form a stacking of buckled honeycomb lattice, which can be grown by cutting-edge
molecular beam epitaxy (MBE) with atomic precision \cite{UedaScience98}. During the growing process,
a single buckled honeycomb layer of B atoms
is replaced by that of nonmagnetic B$'$ atoms, where the element B$'$ is chosen
conjugate to B in order to form a ${\rm d}^8$ configuration.
For B$'$-d electrons on the single buckled honeycomb lattice,
intrinsic SOC becomes sizable, a uniform electric field induces a
staggered electric potential for the two sublattices, and the G-type
AFM order on B atoms on the two sides provides an AFM exchange field.  The material
design makes the magnetic field of pure exchange character, which avoids possible stray
field from permanent ferromagnet.
We have checked successfully our
idea by performing first principles calculations for several materials. In transition metal
perovskites we found intrinsic SOC of several tens of meV, which is larger than that in
silicene in magnitude by one order,
and makes the new topological state available at room temperature.

\vspace{3mm} \textit{Effective model and phase diagram.---}
Let us illustrate our scheme based on the four-band Hamiltonian for
electrons on a buckled honeycomb lattice under an AFM exchange field
and a uniform electric field perpendicular to the plane \cite{note2}

\begin{equation}
H=\sigma_xk_x +\tau_z\sigma_y k_y+\lambda \sigma_z\tau_z s_z
+ V\sigma_z+M\sigma_z s_z,
\end{equation}
around the two inequivalent K and K$'$ points where the $z$ axis is taken
along the [111] direction of peroskite structure;
$\sigma$'s are Pauli matrices for sublattice, and $\tau_z=\pm 1$ and $s_z=\pm 1$
are binary degrees of freedom referring to valley and spin.
The first two terms come from the nearest neighbor hopping same as that in a pristine graphene
\cite{note2}; the third term is for the intrinsic SOC with a positive
coupling coefficient $\lambda>0$ by definition, which can be large
due to the heavier host transition metal atom and the buckled structure;
the fourth term is induced by a uniform electric potential on the buckled structure,
and the last term is the AFM exchange field. The chemical potential is set
to zero, and all energies are measured in units of the nearest neighbor hopping integral
in the present work. As will be shown later, this model Hamiltonian describes the low-energy physics of
transition metal oxides with perovskite structure grown along [111] direction.

\begin{figure}
\psfig{figure=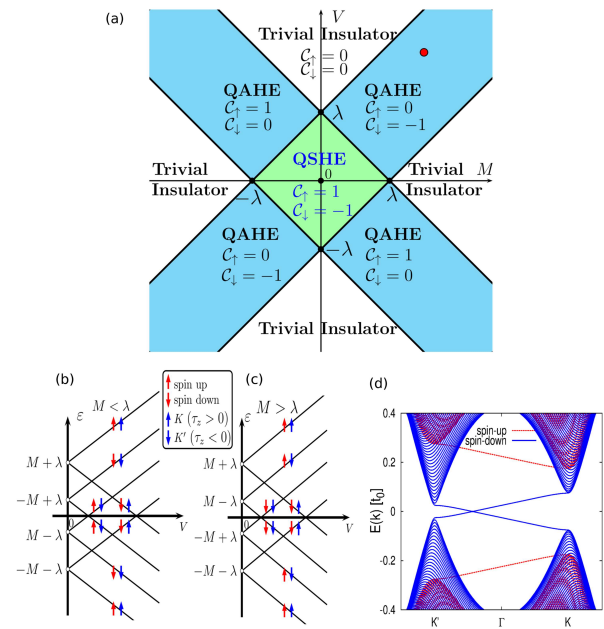,width=8cm}
\caption{(Colour online)
(a) Phase diagram for staggered electric potential $V$ and antiferromagnetic field $M$,
with presumed positive intrinsic SOC $\lambda>0$;
(b) and (c) schematic diagram for energy level crossing upon electric
field tuning for $M<\lambda$ and $M>\lambda$;
(d) band structure for $N=100$ zigzag ribbon of honeycomb lattice with $V=0.25$, $M=0.2$ and $\lambda=0.1$
(red dot in (a)).}
\label{fig:2}
\end{figure}

We reveal behaviors of the system below upon tuning the electric potential
$V$ while the $M$ field and $\lambda$ are fixed, corresponding directly with
our material design and experimental manipulation.
For $0<M<\lambda$, the gaps are topologically nontrivial for both
spin-up and spin-down channels for $V=0$, characterizing the QSHE state (see figure 1(c)).
As illustrated in figure 2(b), when $V$ increases, the gap for spin-up
channel shrinks, closes and reopens at K$'$ point ($\tau_z$=-1) for $V=\lambda-M$.
This quantum phase transition turns the spin-up channel to a trivial state, while the spin-down channel remains
nontrivial. This insulating state exhibiting nontrivial gap in only one
spin channel is different from other topological states such as QSHE and QAHE, as
discussed by Ezawa \cite{EZAWAPRL13}.
When $V$ increases further, the gap for spin-down channel shrinks, closes and
reopens at K point ($\tau_z$=1) for $V=\lambda+M$, which makes the spin-down channel
trivial as well, and the system transforms to a trivial insulating state via
a second quantum phase transition.

For $M>\lambda$, the system exhibits trivial gaps in both spin-up and -down channels
for $V=0$ (see figure 1(c)). As seen in figure 2(c), when $V$ increases, the gap for spin-down channel
shrinks, closes
and reopens at $\tau_z=-1$ for $V=M-\lambda$, which brings the spin-down channel into a topological
state. When $V$ increases further, the gap in spin-down channel shrinks,
closes and reopens at $\tau_z=1$ for $V=\lambda+M$, which drives the system back to a trivial insulator state.

Based on the above analysis we can compose the full phase diagram as displayed in figure 2(a)
by symmetry consideration.
The new topological state is realized when the
absolute values of the three fields $|V|$ and $|M|$
and $|\lambda|$ form a triangle, which indicates clearly that the intervening
among the three degrees of freedom, $\sigma_z$, $\tau_z$ and $s_z$ is crucial for
the new topological state.
A typical band structure for zigzag ribbon of the buckled honeycomb lattice is
displayed in figure 2(d); there are two edge states crossing fermi level which are localized
at the opposite edges (see figure 1(e)).
The present phase diagam is similar to the one by Ezawa \cite{EZAWAPRL13}, except for that
the states with same Chern numbers lie in diagonal directions in the present one, while
locating in same half spaces in that by Ezawa \cite{EZAWAPRL13}, due to the difference of controlling fields
in the two Hamiltonians.

\begin{figure}
\psfig{figure=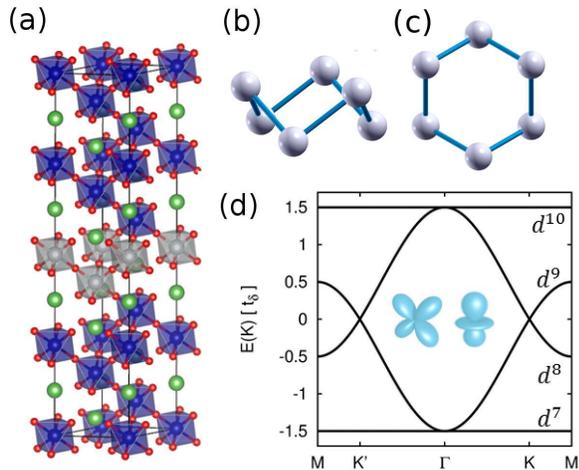,width=8cm}
\caption{(Colour online)
(a) A 6-Layer ABB$'$X perovskite with one layer of B (blue) replaced by
B$'$ (grey). (b) and (c) Side and top views of the buckled honeycomb lattice
formed by B$'$ cations. (d) Noninteracting tight-binding band structure of the e$_{\rm g}$ system,
where each band is labeled with the corresponding electron configuration
if it is occupied. Inset shows the two e$_{\rm g}$ orbits of B$'$. }
\label{fig:3}
\end{figure}

\vspace{3mm} \textit{Material realization.---}
In perovskite ABO$_3$ material shown in figure 3(a),
A and B form two penetrating simple cubic lattices. The octahedron cage formed by
six oxygens around the transition metal B generates the doublet e$_{\rm g}$ and triplet t$_{\rm 2g}$ orbits
of the d electrons. In the [111] direction, a simple cubic
structure can be viewed as a stacking of buckled honeycomb lattices as shown in figures 3(b) and (c).
In the new coordinate, intrinsic
SOC appears between the two e$_{\rm g}$ orbits. In regard of material synthesis, recent developments in
laser molecular beam epitaxy (MBE) technique permit one to grow perovskite structure along [111] direction with
atomic precision \cite{UedaScience98}. All these make the d electrons in perovskite structure
a very hopeful platform for realizing topological states, as illustrated first by
Xiao et al. \cite{XiaoNatComm11}.

\begin{figure}
\psfig{figure=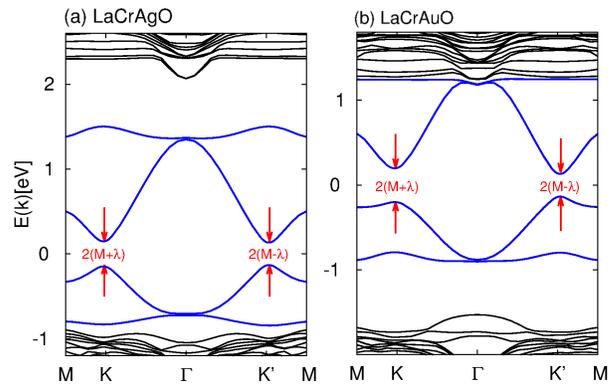,width=8cm}
\caption{(Colour online)
Band structures based on first-principles calculation on LaCrAgO and LaCrAuO with a super unit cell
(A$_2$B$'$$_2$O$_6$)/(A$_2$B$_2$O$_6$)$_{5}$ along the [111] direction.}
\label{fig:4}
\end{figure}

Many transition metal oxides with perovskite structure are known as wide-gap antiferromagnetic insulator.
A subgroup of these materials presume the so-called G-type AMF order where spins
align antiferromagnetically in the two sublattices of the buckled honeycomb lattice.
This is the playground where we perform material design and field manipulation to realize
topological state with our scheme.
We replace one buckled layer of B by a nonmagnetic element B$'$, and obtain B$'$-d electrons on a single layer of buckled
honeycomb lattice, which feel AFM exchange field from
the G-type AMF order of the host material ABO$_3$ \cite{note1}.

\begin{table}
\caption{Parameters for AFM order and SOC fit from GGA+U+SOC calculation.
For KNiInF and KNiTlF the electronic configurations of In$^{2+}$ and
Tl$^{2+}$ are $5s^1$ and $6s^1$ . }
\begin{tabular}  {c c c c c c }
  \hline
  \hline
   field [meV]  & LaCrAgO & LaCrAuO & LaFeAgO & LaFeAuO   \\ \hline
  $M$   & 141 & 166 & 541 & 467  \\
  $\lambda$   & 7.30 & 32.91  & 7.31 & 33.52    \\
  \hline
  \hline
  KNiPdF & KNiPtF & KNiInF$^{\ast}$ & KNiTlF$^{\ast}$ & \\ \hline
   625 & 504 & 290 & 235 &\\
   11.38 & 33.40 & 5.05 & 18.58&\\
     \hline
  \hline
\end{tabular}
\end{table}

There is a gap (10Dq) between the e$_{\rm g}$ and t$_{\rm 2g}$ orbits, which drops t$_{\rm 2g}$ orbits from the low-energy
physics. In each unit cell, two B$'$ atoms contribute totally four e$_{\rm g}$ orbits.
The band structure for these four orbits is shown in figure 3(d) based on a tight-binding Hamiltonian
with the hopping integrals given by Slater-Koster formula. One finds two
flat bands and two dispersive bands crossing each other. Xiao et al. \cite{XiaoNatComm11,RueggPRB11,YangPRB11}
focus on the flat band and developed an interesting scenario for possible topological state
based on strong electron correlations by using ${\rm d}^7 ({\rm t}_{\rm 2g}^6{\rm e}_{\rm g}^1)$ configuration.
Here we concentrate on the two dispersive bands by taking ${\rm d}^8({\rm t}_{\rm 2g}^6{\rm e}_{\rm g}^2)$
systems for which the fermi level is around their crossing point.

In order to verify the above picture, we have performed first principles calculation
on the host material LaCrO$_3$ with one layer of La$_2$Ag$_2$O$_6$ or La$_2$Au$_2$O$_6$
inserted, which realizes a
${\rm d}^8({\rm t}_{\rm 2g}^6{\rm e}_{\rm g}^2)$
system on the buckled honeycomb lattice. The super cell
used in the calculations contains 6 layers with a configuration of
(La$_2$B$'$$_2$O$_6$)/(La$_2$B$_2$O$_6$)$_{5}$. The calculations are performed by using
density functional theory(DFT) within the generalized gradient
approximation (GGA) approach in the parametrization of Perdew, Burke and
Ehzorhof (PBE) \cite{PBE} for exchage-correlation as implemented in the
Vienna Ab Initio Simulation Package (VASP) \cite{VASP}. The bulk lattice
constant is set to $3.85{\AA}$ for LaCrO$_3$ and atoms are allowed to
relax inside the super cell. The on-site Coulomb interactions of the $3d$
electrons of Cr are treated by Dudarev' method with an effective U value
\cite{Dudarev}, U$_{eff}$=U-J=5.4eV. We have checked that our main results
are valid for a large window of the U values. The momentum space-mesh is
$6\times6\times1$ and the cutoff energy is 440eV.

The band structures of LaCrAgO and LaCrAuO obtained from GGA+U+SOC calculations are displayed
in figures 4(a) and (b). First of all, the two flat bands and two dispersive bands
close to the fermi level in figure 3(d) are realized in the two materials. The two
gaps at K and K$'$, which were absent in figure 3(b) for tight-binding model,
are induced by the AFM exchange field from LaCrO$_3$ and the SOC. From the
two gaps at K and K$'$, we figure out $M=0.141$eV and $\lambda=7.3$meV for LaCrAgO,
and $M=0.166$eV and $\lambda=32.91$meV for LaCrAuO.
We have also calculated several other possible materials and the results are
summarized in TABLE 1. As the maximal
nontrivial gap is given by $2\lambda$, the results from first principles calculations
suggest that a topological material may be realized at room temperature.
Typical electric field to realize the
new topological state in the order of 0.1V/$\AA$.

\begin{figure}
\psfig{figure=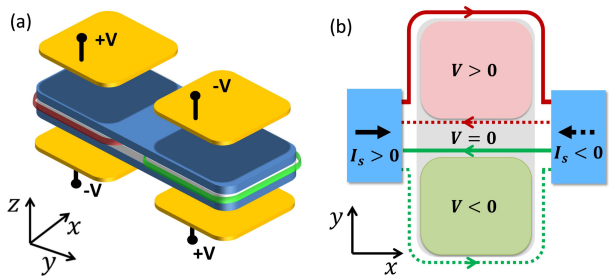,width=8cm}
\caption{(Colour online)
Device for a "composite QHSE" insulator formed by two copies of the new topological state.}
\label{fig:5}
\end{figure}

\vspace{3mm} \textit{Possible applications.---}
This new topological state available probably at room temperature
is ideal for spintronics applications, since it provides a spin-polarized quantized edge
current, robust to both nonmagnetic and magnetic defects, with the spin polarization
tunable by inverting electric field.

Using two copies of the new topological state, one can realize
a "composite QSHE insulator". As shown in figure 5(a), we fabricate two pairs of electrodes
(golden) with opposite bias voltages. The two patches exhibit states with
opposite chiralities and spins.
Spin Hall conductance can be measured based on this device. As shown in figure 5(b),
when a positive spin current $I_s$ is
injected from the left electrode, the out-going spin-up current flows from
the left electrode along the top edge of upper patch (upturned red solid
curve), while the in-going spin-down current flows into the left electrode along the straight line
between the two electrodes (solid green line), which results in zero charge current.
Because charge current flows only on the top edge and
cumulates charges there, a Hall voltage drop can be detected between the
top and bottom edges of the device.
As compared with the QSHE realized before,
even a time-reversal-symmetry-broken perturbation, e.g., a magnetic field
or magnetic impurities cannot backscatter the counter charge currents with spin up
and down at the center of the device, since the
charge currents are now distributed in different patches as shown in figure 5(b).

\vspace{3mm} \textit{Discussions.---}
The present topological state is
robust against Rashba SOC \cite{QiaoPRB10,EZAWAPRL12}, which, although
small, should be present since the electric field breaks the inversion symmetry with
respect to the honeycomb lattice plane. Actually, it is known that in graphene
the nontrivial band gap closes for $\lambda_R>\lambda$ and the
QSHE will be destroyed\cite{KaneMele05} due to the mixing of the
two spin channels by Rashba SOC. In the new
topological state with for example $V=M>0$, a gap of $2(2M-\lambda)$ is
opened in spin-up channel whereas the gap in the spin-down channel is
$2\lambda$. Since $M\gg \lambda$ as shown in Table 1, even
a Rashba SOC stronger than the intrinsic one $\lambda_R\gtrsim \lambda$
cannot close the gap by mixing the two spin channels.

\vspace{3mm} \textit{Note.---} About finishing our manuscript, we became aware of a
recent work by Ezawa arXiv.1301.0971, addressing the same physics.
In his work, the staggered magnetic field is generated by two ferromagnets
on the two sides of silicene layer.

\vspace{5mm}
\noindent\textit{Acknowledgements.---}
X.H. is grateful to Masashi Tachiki and Seiji Miyashita for helpful discussions.
This work was supported by WPI
Initiative on Materials Nanoarchitectonics, MEXT of Japan, and
Grants-in-Aid for Scientific Research (No.22540377), JSPS, and partially
by CREST, JST. Q.F.L. is also supported by NSFC under grants 10904092.


\begin{thebibliography}{99}
\bibitem{note2}
  For $d^8$ system where the electronic structures involves two degenerated $e_{g}$ orbits, one can still obtain an
  effective $H_0$ by reducing a 4-band hamiltonian into a 2-band one using $k\cdot p$ perturbation theory.
\bibitem{note1}
  Actually there are two magnetic configurations which have similar energies.  From our first principle calculation, we
  found the one which induces the desired AFM magnetizaiton in the AB'O3 layer always has a lower energy.
\end{thebibliography}
\end{document}